\documentclass[twoside]{dis04}
\def\runauthor{A. Szczurek and M. {\L}uszczak}
\def\shorttitle{$c-\bar c$ correlations}
\begin{document}

\title{CHARM-ANTICHARM KINEMATICAL CORRELATIONS \\
IN PHOTON-PROTON SCATTERING \\
}

\author{Antoni Szczurek$^{1,2}$, Marta {\L}uszczak$^{2}$}

\address{$^{1}$Institute of Nuclear Physics\\
ul. Radzikowskiego 152, PL-31342 Cracow, Poland\\
and \\
$^2$University of Rzesz\'ow\\
PL-35-959 Rzesz\'ow, Poland\\
E-mail: antoni.szczurek@ifj.edu.pl }

\maketitle

\abstracts{
$c-\bar c$ correlations are calculated
in the $k_t$-factorization approach. Different unintegrated
gluon distributions (uGDF) from the literature are used.
The results are compared with recent results of
the FOCUS collaboration. The recently developed CCFM uGDF gives
a good description of the data.
Predictions for the HERA energies are presented.
}

\section{Introduction}

In recent years a lot of activity was devoted to the description of
the photon-proton total cross section in terms of
the unintegrated gluon distribution functions (uGDF)
(see e.g.\cite{small-x-02,small-x-03} and references therein).
In some of the analyses also inclusive charm quark (or meson)
were considered \cite{QQbar}. Although the formalism of uGDF is well
suited for studying more exclusive observables,
only very few selected cases were considered in the literature.
A special example is azimuthal jet-jet correlations
in photon-proton scattering \cite{SNSS}.
Recently the FOCUS collaboration at Fermilab provided precise
data for $c-\bar c$ correlations \cite{FOCUS}.
We analyze the $c-\bar c$ correlations
in terms of uGDF available in the literature.
While the total cross section depends on small values of x,
the high-$p_t$ jets and/or heavy quark production test
gluon distributions at somewhat larger x.
Only some approches are applicable in this region.
In particular, we test results obtained with recently developed
CCFM unintegrated parton distributions. This presentation is based
on \cite{LS04} where more details can be found.

\section{Charm-anticharm correlations}

The total cross section for quark-antiquark production
in the reaction $\gamma + p \to Q + \bar Q + X$ can be written as
\cite{SNSS}
\begin{equation}
\sigma^{\gamma p \to Q \bar Q}(W) =
\int d \phi \int dp_{1,t}^2 \int dp_{2,t}^2 \int dz \;
\frac{f_g(x_g,\kappa^2)}{\kappa^4} \cdot
\tilde{\sigma}(W,\vec{p}_{1,t},\vec{p}_{2,t},z) \; .
\label{master_formula}
\end{equation}
In the formula above $f_g(x,\kappa^2)$ is the unintegrated gluon
distribution.
The gluon transverse momentum is related to the quark/antiquark
transverse momenta $\vec{p}_{1,t}$ and $\vec{p}_{2,t}$ as:
\begin{equation}
\kappa^2 = p_{1,t}^2 + p_{2,t}^2 + 2 p_{1,t} p_{2,t} cos\phi \; .
\end{equation}
Above we have introduced:
\begin{eqnarray}
\tilde{\sigma}(W,\vec{p}_{1,t},\vec{p}_{2,t},z)
= \frac{\alpha_{em}}{2} \; e_Q^2 \; \alpha_s(l^2) \nonumber \\
\left\{
   [ z^2+(1-z)^2 ] \;
  \Bigg\vert \; \frac{\vec{p}_{1,t}}{p_{1,t}^2 + m_Q^2} +
     \frac{\vec{p}_{2,t}}{p_{2,t}^2 + m_Q^2} \;
  \Bigg\vert ^2
   \; + \;
   m_Q^2 \left(
   \frac{1}{p_{1,t}^2 + m_Q^2} +
   \frac{1}{p_{2,t}^2 + m_Q^2} +
      \right)^2 
\right\}
\label{aux}
\end{eqnarray}
The unintegrated gluon distribution $f_g$ is evaluted at
\begin{equation}
x_g = \frac{M_t^2}{W^2} \; ,
\label{x_g}
\end{equation}
where
\begin{equation}
M_t^2 = \frac{p_{1,t}^2 + m_Q^2}{z} + \frac{p_{2,t}^2 + m_Q^2}{1-z}
\; .
\label{invariant_mass}
\end{equation}

The basic ingredient of our approach are unintegrated gluon
distributions. Different models of uGDF have been proposed
in the literature (see for instance \cite{small-x-02,small-x-03}).
The main effort has been concentrated
on the small-x region. While the total cross section is the genuine
small-x phenomenon (x $<$ 10$^{-3}$), the production of
charm and bottom quarks samples rather the intermediate-x region
(x$\sim$ 10$^{-2}$ - 10$^{-1}$) even at the largest available
energies at HERA. It is not obvious if the methods used are
appropriate for the intermediate values of x.
In the present approach we shall present results for a few
selected gluon distributions from the literature.
For illustration we shall consider a simple BFKL \cite{BFKL},
saturation model used to study HERA photon-proton total cross
sections \cite{GBW_glue} (GBW),
saturation model used recently to calculate particle production
in hadron-hadron collisions \cite{KL01} (KL). These three model
approaches are expected to be valid for low values of x.
At somewhat larger values of x all these models are expected to break. 
As in Ref.\cite{pp_pions} we shall try to extend
the applicability of the model by multiplying the model
distributions by a phenomenological factor $(1-x)^n$. 

\begin{figure}[htb] 
    \includegraphics[width=5.0cm]{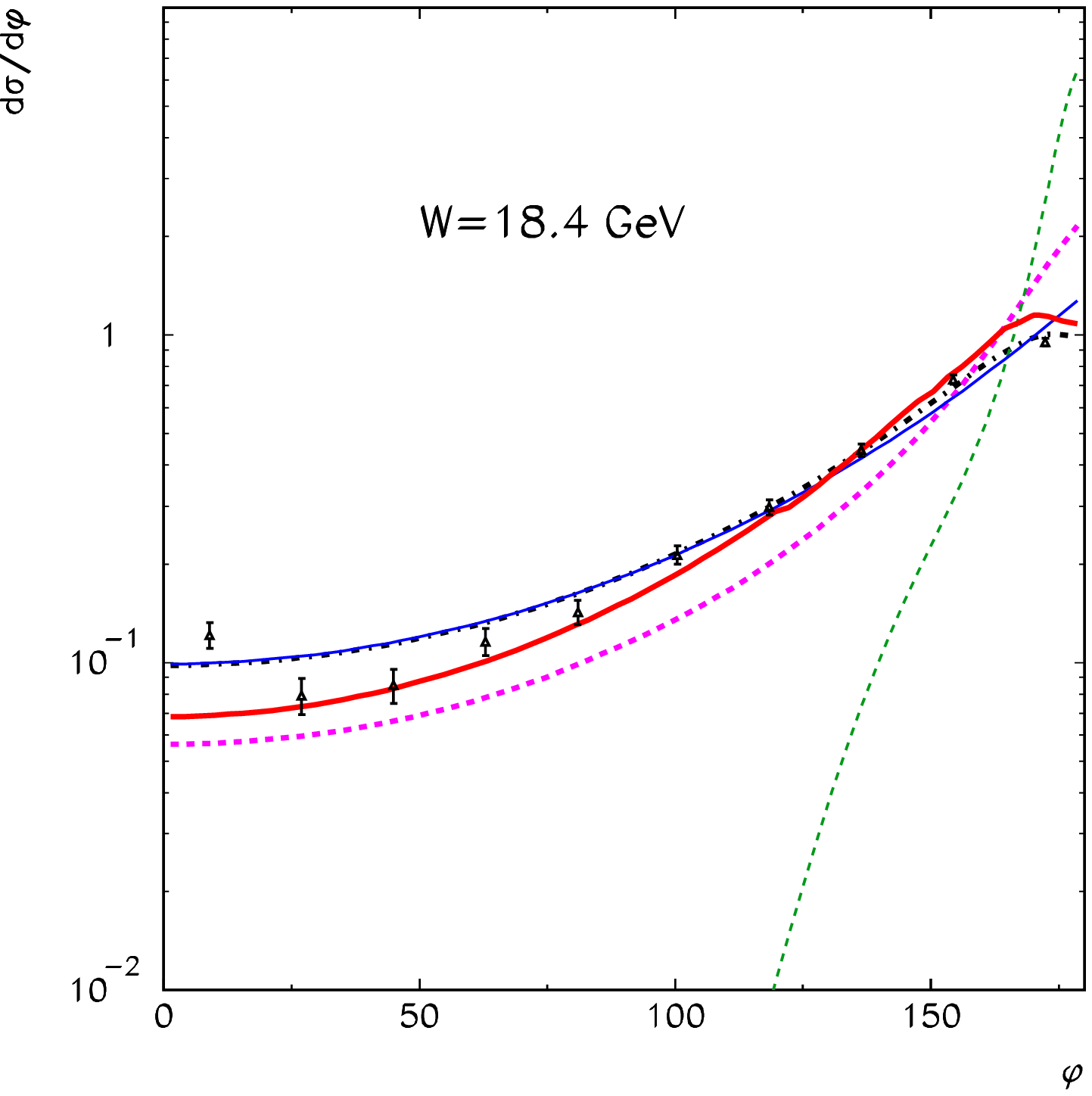}
    \includegraphics[width=5.0cm]{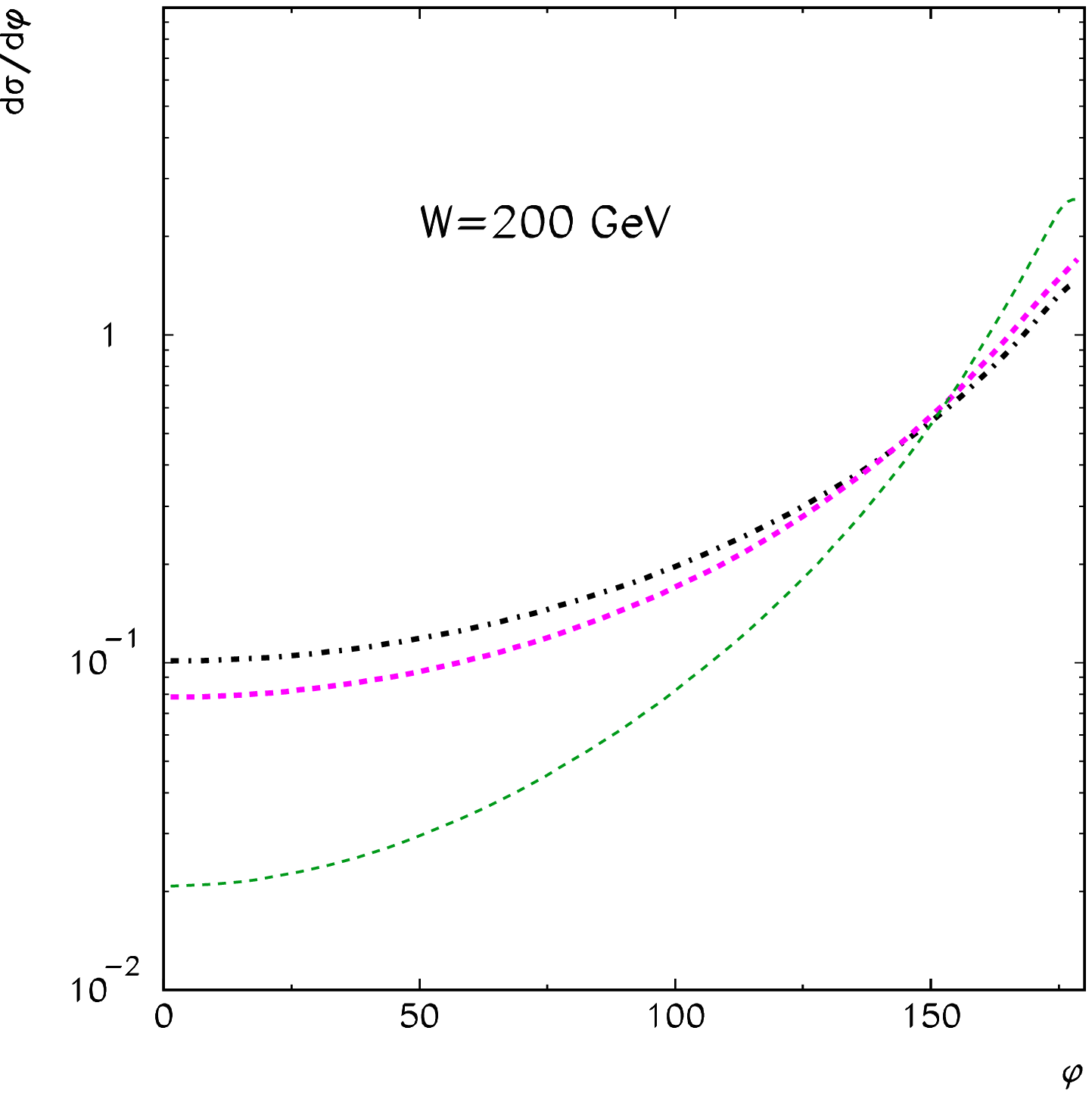}
\caption[*]{
Azimuthal correlations between $c$ and $\bar c$.
The theoretical results are compared to the recent results
from \cite{FOCUS} (fully reconstructed pairs).
\label{fig_phi}
}
\end{figure}

The azimuthal correlation functions $w(\phi)$
defined as:
\begin{equation}
w(\phi) =
 \int dp_{1,t}^2 \int dp_{2,t}^2 \int dz \;
\frac{f_g(x_g,\kappa^2)}{\kappa^4} \cdot
\tilde{\sigma}(W,\vec{p}_{1,t},\vec{p}_{2,t},z) \; .
\end{equation}
and normalized to unity
for two energies of $W$ = 18.4 GeV (FOCUS)
and $W$ = 200 GeV (HERA) are shown in Fig.\ref{fig_phi}.
The GBW-glue (thin dashed) gives too strong back-to-back correlations
for the lower energy. Another saturation model (KL, \cite{KL01})
provides more angular decorrelation, in better agreement with
the experimental data. The BFKL-glue (dash-dotted) provides very
good description of the data. The same is true for the
CCFM-glue (thick solid) and resummation-glue (thin solid).
The latter two models are more adequate for the lower energy.
The renormalized azimuthal correlation function
for BFKL, GBW and KL models are almost independent of the power
$n$ in extrapolating to larger values of $x_g$.
In calculating the cross section with the CCFM uGDF for simplicity
we have fixed the scale for $\mu^2$ = 4 $m_c^2$.
In the present calculations we have used exponential form factor
with $b_e$ = 0.5 GeV$^{-1}$ (see \cite{KS04}).
For comparison in panel (b) we present predictions for $W$ = 200 GeV.
Except of the GBW model, there is only a small increase of
decorrelation when going from the lower fixed-order energy region
to the higher collider-energy region.

\begin{figure}[htb] 
    \includegraphics[width=5.0cm]{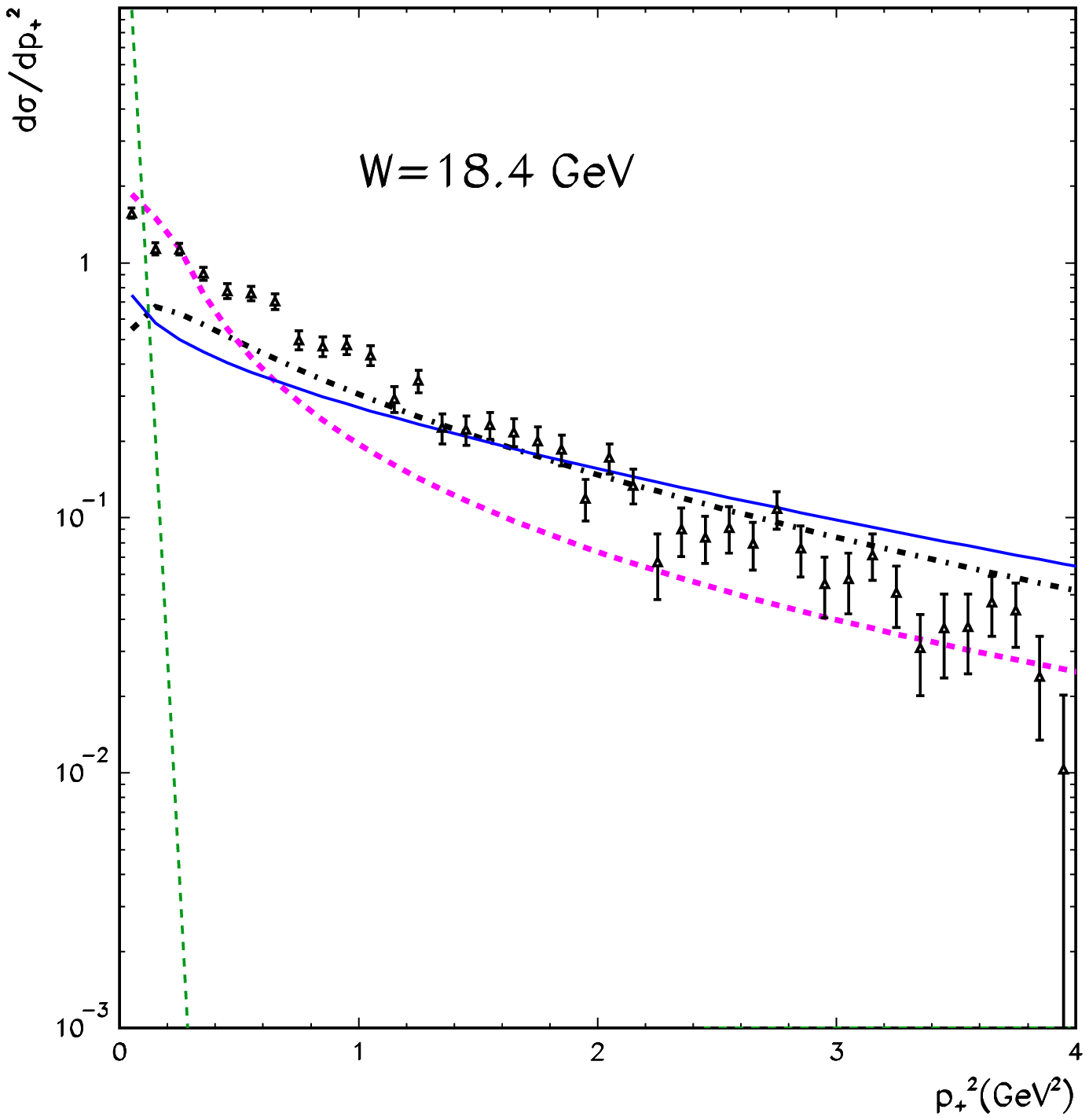}
    \includegraphics[width=5.0cm]{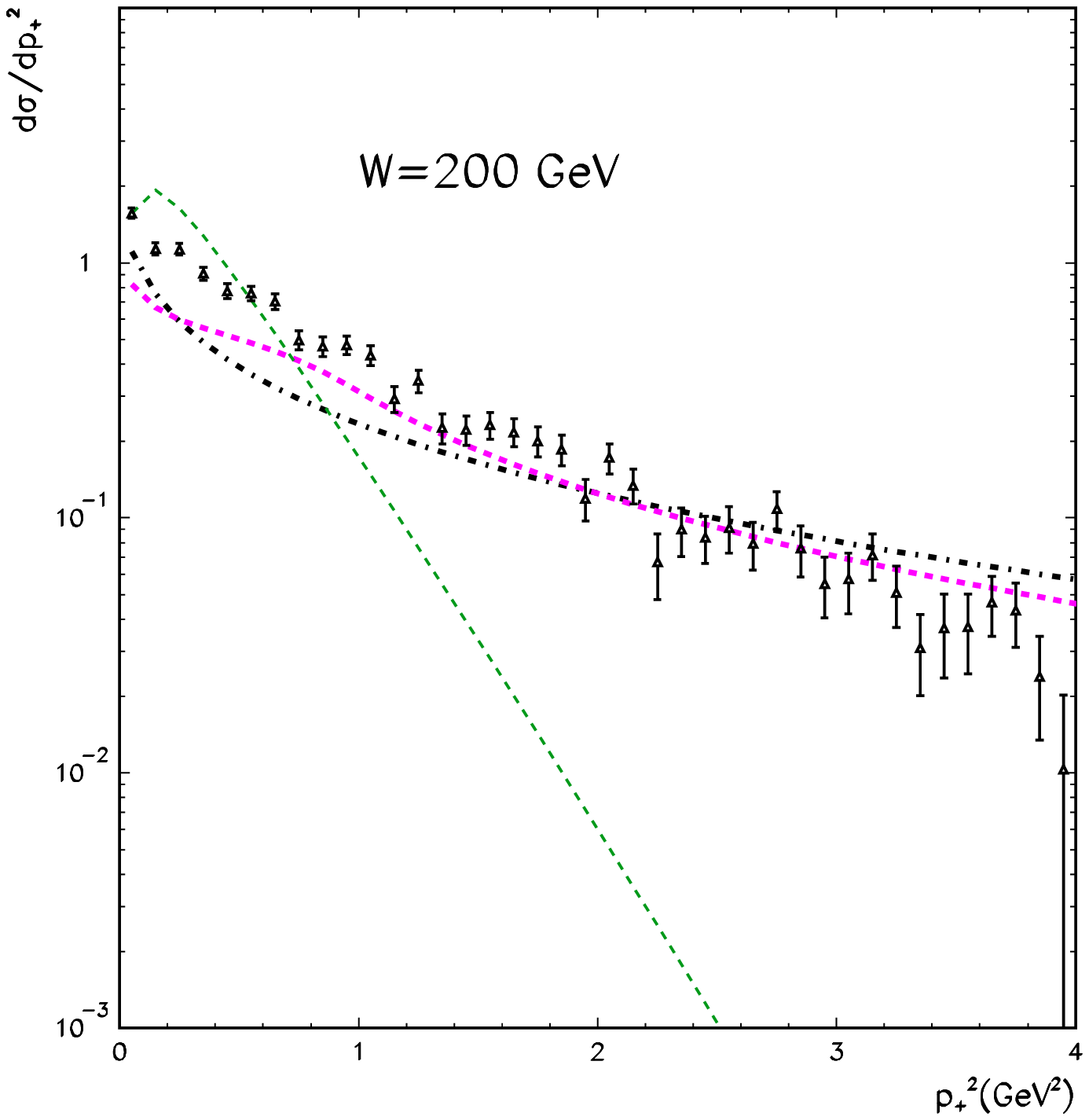}
\caption[*]{
$p_+^2$ distribution of $c - \bar c$. 
The theoretical results are compared to the recent results
from \cite{FOCUS} (fully reconstruced pairs).
The notation is the same as in Fig.\ref{fig_phi}.
\label{fig_psum2}
}
\end{figure}

In the LO collinear approach the transverse momenta
of two jets add up to zero. It is not the case for our LO
$k_t$-factorization approach. In Fig.\ref{fig_psum2} we present
normalized to unity distribution in $p_{+}^2$, where
$\vec{p}_{+} = \vec{p}_1 + \vec{p}_2$. Due to momentum conservation,
in our approach the sum of
transverse momenta is directly equal to transverse momentum of
the gluon ($\vec{p}_{+} = \vec{\kappa}$). This means that
the distribution in $p_{+}^2$ directly probes the transverse
momentum distribution of gluons.
The CCFM gluon distribution gives the best description of
the FOCUS data \cite{FOCUS}. We expect that this
approach is suitable for $x > 0.01$. Although the other models
give also reasonable description of the FOCUS data, one should
remember that their application for the low energy data is
somewhat unsure.

\begin{figure}[htb] 
    \includegraphics[width=5.0cm]{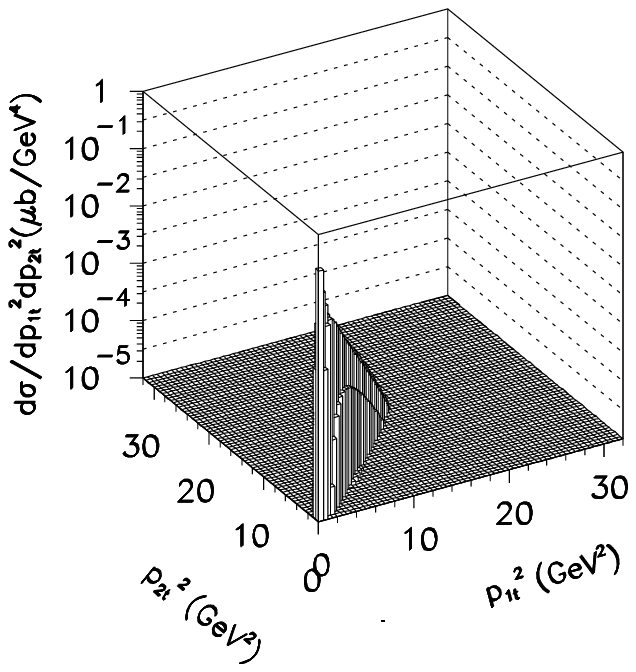}
    \includegraphics[width=5.0cm]{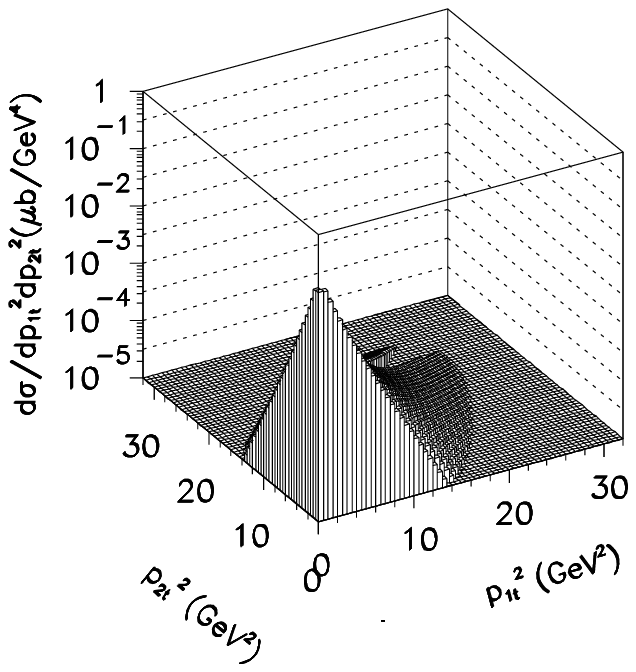}
\caption[*]{
The two-dimensional maps $w(p_{1,t}^2,p_{2,t}^2)$
for W = 18.4 GeV: a) GBW and b) CCFM.
\label{fig_maps}
}
\end{figure}

Not only azimuthal correlations are interesting.
The formula (\ref{master_formula}) can be rewritten in the form
\begin{equation}
\sigma^{\gamma p \to Q \bar Q}(W) =
\int dp_{1,t}^2 \int dp_{2,t}^2 \;
w(p_{1,t}^2,p_{2,t}^2;W)  \; ,
\label{p1t2_p2t2_correlations}
\end{equation}
where the two-dimensional correlation function
\begin{equation}
w(p_{1,t}^2,p_{2,t}^2;W) =
\int d \phi \; dz  \;
\frac{f_g(x_g,\kappa^2)}{\kappa^4} \cdot
\tilde{\sigma}(W,\vec{p}_{1,t},\vec{p}_{2,t},z) \; .
\label{w_p1t2_p2t2}
\end{equation}
In Fig.\ref{fig_maps} we present two examples of
$w(p_{1,t}^2, p_{2,t}^2)$ at W=18.4 GeV for different uGDF.
As in Ref.\cite{szczurek02} for GBW we have used
the power n=7 in the extrapolating formula.
While normalization is slightly dependent on the value of the power,
the shape of the two-dimensional map is almost the same.
The maps for different uGDF differ in details.
The distribution for the GBW gluon distribution is concentrated along
the diagonal $p_{1,t}^2 = p_{2,t}^2$ and in this respect resembles
the familiar collinear LO result.
The CCFM distribution has a sizeable strength at the
phase-space borders for
$p_{1,t}^2 \approx$ 0 or $p_{2,t}^2 \approx$ 0.
Experimental studies of such maps could open a possibility
to test models of uGDF in a more detailed differential way.
In principle, such studies will be possible with HERA II runs
at DESY.

\section{Summary}

We have shown that analysis of kinematical correlations
of charm quarks opens a new possibility to verify models of uGDF.
The recently measured data of the FOCUS collaboration at Fermilab
allows to study the unintegrated gluon distribution in the
intermediate-x region. The recently developed unintegrated
gluon (parton) distributions which fulfil the CCFM equation
describe the data fairly well. Many models of uGDF from the literature
are constructed rather for small values of x and its application
in the region of somewhat larger x (x $>$ 0.05) is questionable.
It can be expected that the correlation data from the HERA II runs
will give a new possibility to verify the different models of
unintegrated gluon distributions in more detailed way.

\end{document}